\documentclass[runningheads]{llncs}
\usepackage{epsfig}
\usepackage{graphicx}
\usepackage{amsmath}
\usepackage{amssymb}
\usepackage{pdfrender}

\usepackage{subfigure}
\usepackage{multirow}
\usepackage{array}
\usepackage{color}
\usepackage{tabu}
\usepackage{longtable}

\usepackage[breaklinks=true,letterpaper=true,colorlinks,bookmarks=false]{hyperref}
\newcommand{\etal}{\textit{et al}. }
\newcommand{\ie}{\textit{i}.\textit{e}., }
\newcommand{\eg}{\textit{e}.\textit{g}., }

\begin{document}
\title{Lesion Segmentation and RECIST Diameter Prediction via Click-driven Attention and Dual-path Connection}
\titlerunning{Lesion Segmentation and RECIST diameter prediction}
%

\author{Youbao Tang\inst{1} \and
Ke Yan\inst{1} \and
Jinzheng Cai\inst{1} \and
Lingyun Huang\inst{2} \and
Guotong Xie\inst{2} \and
Jing Xiao\inst{2} \and
Jingjing Lu\inst{3} \and
Gigin Lin\inst{4} \and
Le Lu\inst{1}}

\institute{PAII Inc., Bathesda, MD, USA \\ \email{tybxiaobao@gmail.com; tiger.lelu@gmail.com} \and Ping An Technology, Shenzhen, PRC \and Beijing United Family Hospital, Beijing, PRC \and Chang Gung Memorial Hospital, Linkou, Taiwan, ROC}

\maketitle              

\begin{abstract}
Measuring lesion size is an important step to assess tumor growth and monitor disease progression and therapy response in oncology image analysis. Although it is tedious and highly time-consuming, radiologists have to work on this task by using RECIST criteria (Response Evaluation Criteria In Solid Tumors) routinely and manually. Even though lesion segmentation may be the more accurate and clinically more valuable means, physicians can not manually segment lesions as now since much more heavy laboring will be required. In this paper, we present a prior-guided dual-path network (PDNet) to segment common types of lesions throughout the whole body and predict their RECIST diameters accurately and automatically. Similar to \cite{tang2020one}, a click guidance from radiologists is the only requirement. There are two key characteristics in PDNet: 1) Learning lesion-specific attention matrices in parallel from the click prior information by the proposed prior encoder, named click-driven attention; 2) Aggregating the extracted multi-scale features comprehensively by introducing top-down and bottom-up connections in the proposed decoder, named dual-path connection. Experiments show the superiority of our proposed PDNet in lesion segmentation and RECIST diameter prediction using the DeepLesion dataset and an external test set. PDNet learns comprehensive and representative deep image features for our tasks and produces more accurate results on both lesion segmentation and RECIST diameter prediction.

\keywords{Weakly-supervised lesion segmentation \and RECIST diameter prediction \and Click-driven attention \and Dual-path connection.}
\end{abstract}
\section{Introduction}
Assessing lesion growth across multiple time points is a major task for radiologists and oncologists. The sizes of lesions are important clinical indicators for monitoring disease progression and therapy response in oncology. A widely-used guideline is RECIST (Response Evaluation Criteria In Solid Tumors)~\cite{eisenhauer2009new}, which requires users to first select an axial slice where the lesion has the largest spatial extent, then measure the longest diameter of the lesion (long axis), followed by its longest perpendicular diameter (short axis). This process is highly tedious and time-consuming. More importantly, it is prone to inconsistency between different observers~\cite{tang2018semi}, even with considerable clinical knowledge. Segmentation masks may be another quantitative and meaningful metric to assess lesion sizes, which is arguably more accurate/precise than RECIST diameters and avoids the subjectivity of selecting long and short axes. However, it is impractical and infeasible for radiologists to manually delineate the contour of every target lesion on a daily basis due to the heavy work load that would require.

Deep learning based computer-aided diagnosis techniques \cite{wang2020knowledge,tang2020automated,tang2020e2net,yan2020learning,cai2020deep,yan2020self,tang2021disentangled,cheng2021scalable} have been extensively studied by researchers, including automatic lesion segmentation. Most existing works focused on tumors of specific types, such as lung nodules~\cite{WANG2017172,jin2018ct}, liver tumors~\cite{li2018h,tang2020e2net}, and lymph nodes~\cite{zhu2020lymph}. However, radiologists often encounter different types of lesions when reading an image. Universal lesion segmentation~\cite{cai2018accurate,tang2018ct,tang2020one,agarwal2020weakly,tang2021ahrnet} and measurement~\cite{tang2018semi,tang2020one} have drawn attentions in recent years, aiming at learning from a large-scale dataset to handle a variety of lesions in one algorithm. These work leverage NIH DeepLesion dataset~\cite{yan2018deeplesion}, which contains the RECIST annotations of over 30K lesions of various types. Among them, \cite{tang2018semi} requires users to draw a box around the lesion to indicate the lesion of interest. It first employs a spatial transform network to normalize the lesion region, then adapts a stacked hourglass network to regress the four endpoints of the RECIST diameters. \cite{tang2020one} requests users to only click a point on or near the lesion, more convenient and efficient than \cite{tang2018semi}. It uses an improved mask R-CNN to detect the lesion region and subsequently performs segmentation and RECIST diameter prediction \cite{tang2020one}. User click information is fed into the model as the input together with the image. This strategy treats lesions with diverse sizes and shapes in the same way, thus may not be optimal at locating the lesion region precisely.

In this paper, we propose a novel framework named prior-guided dual-path network (PDNet). Following \cite{tang2020one}, given a 2D computed tomography (CT) slice and a click guidance in a lesion region, our goal is to segment the lesion and predict its RECIST diameters automatically and reliably. To achieve this goal, we adopt a two-stage framework. The first stage extracts the lesion of interest (LOI) by segmentation rather than detection in \cite{tang2020one}, since sometimes the detection results do not cover the lesions that we click in. The second stage obtains the lesion segmentation and RECIST diameter prediction results from the extracted LOI. We propose a novel prior encoder to encode the click prior information into attention maps, which can deal with considerable size and shape variations of the lesions. We also design a scale-aware attention block with dual-path connection to improve the decoder. PDNet is evaluated on manually-labeled lesion masks and RECIST diameters in DeepLesion dataset~\cite{yan2018deeplesion}. To prove the generalizability of our method, we additionally collected an external test set from 6 public lesion datasets of 5 organs. Experimental results show that PDNet outperforms the previous state-of-the-art method~\cite{tang2020one} and a strong baseline nnUNet~\cite{isensee2018nnu} on two test sets, for both lesion segmentation and RECIST diameter prediction tasks.

\section{Methodology}

\begin{figure*}[t!]
  \centering
  \includegraphics[width=0.99\linewidth]{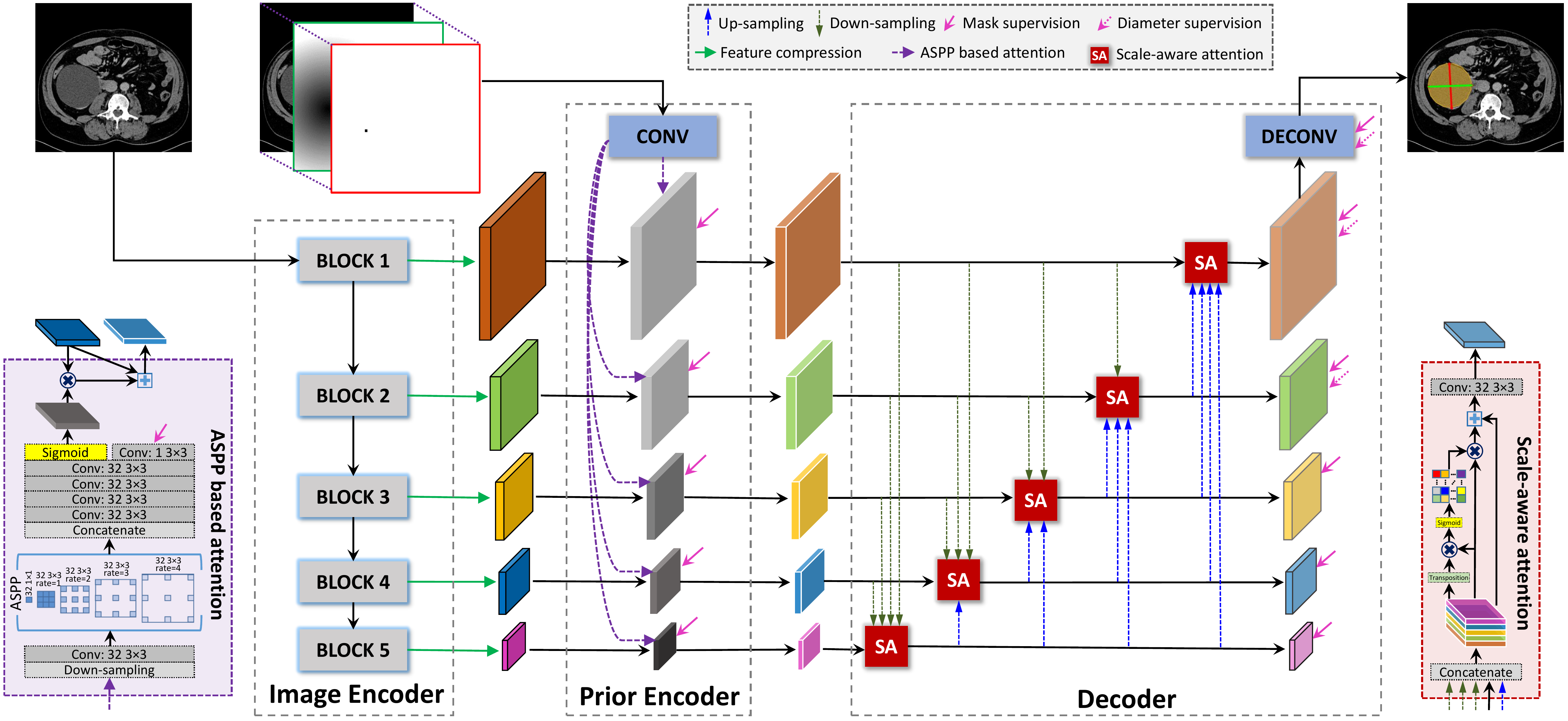}
  \caption{Overview of the proposed prior-guided dual-path network (PDNet).}
  \label{fig:framework}
\end{figure*}

Our framework includes two stages. The first stage extracts the lesion of interest (LOI) by segmentation; the second stage performs lesion segmentation and RECIST diameter prediction from the extracted LOI. A prior-guided dual-path network (PDNet) is proposed for the tasks at both stages. Fig. \ref{fig:framework} shows the overview of the proposed PDNet. It consists of three components: an image encoder, a prior encoder with click-driven attention, and a decoder with dual-path connection.

\textbf{Image Encoder:}
The image encoder aims to extract highly discriminative features from an input CT image. Also, representing features at multiple scales is of great importance for our tasks. Recently, Zhang \etal \cite{zhang2020resnest} present a split-attention block and stack several such blocks in ResNet style \cite{He_2016_CVPR} to create a split-attention network, named ResNeSt. ResNeSt is able to capture cross-channel feature correlations by combining the channel-wise attention with multi-path network layout. Extensive experiments demonstrate that it universally improves the learned feature representations to boost performance across numerous vision tasks. Therefore, this work utilizes ResNeSt-50 \cite{zhang2020resnest} as backbone to extract highly discriminative multi-scale features in the image encoder. As shown in Fig. \ref{fig:framework}, ResNeSt-50 has five blocks that output multi-scale features with different channels. To relieve the computation burden, they are compressed to 32 channels using a convolutional layer with 32 $3 \times 3$ kernels.

\textbf{Prior Encoder with Click-driven Attention:}
Given a click guidance, a click image and a distance transform image are generated and considered as prior information following \cite{tang2020one}, as shown in Fig. \ref{fig:framework}. In \cite{tang2020one}, the prior information is integrated into the model by directly treating it as input for feature extraction. We argue that the representation ability of features extracted from image encoder may be weaken by this strategy. That is because the sizes and shapes of different lesions are highly diverse, but their prior information generated using this strategy are the same. To avoid this, we separately build a prior encoder (PE) with click-driven attention, which is able to learn lesion-specific attention matrices by effectively exploring the click prior information. With them, the representation ability of the extracted multi-scale features from image encoder will be enhanced to improve the performance of our tasks. As shown in Fig. \ref{fig:framework}, the prior encoder takes as input the compressed multi-scale features and a 3-channel image (the original CT image, the click image, and the distance transform image), and outputs attention enhanced multi-scale features. The prior encoder includes five atrous spatial pyramid pooling (ASPP) \cite{chen2018deeplab} based attention modules and a convolutional layer with 32 3$\times$3 kernels and a stride of 2. The detailed structure of ASPP based attention module can be found from the purple box in Fig. \ref{fig:framework}, where 5 side outputs (the pink solid arrows) are added to introduce the deep mask supervision to learn the attention matrices.

\textbf{Decoder with Dual-path Connection:}
It is known that the low-level scale features focus on fine-grained lesion parts (\eg edges) but are short of global contextual information, while the high-level scale features are capable of segmenting the entire lesion regions coarsely but at the cost of losing some detailed information. With this inspiration, unlike UNet \cite{ronneberger2015u} where the decoder only considers current scale features and its neighbouring high-level scale features gradually, we build a new decoder that can aggregate the attention enhanced multi-scale features more comprehensively. Specifically, each scale features are reasonably interacted with all lower-level and higher-level scale features in the decoder, which is accomplished by using dual-path connection (\ie top-down connection and bottom-up connection). The top-down connection (T2D) adopts a bilinear interpolation operation on the high-level scale features for up-sampling followed by a convolutional layer with 32 3$\times$3 kernels for smoothing. The bottom-up connection (B2U) performs a convolution operation with 32 3$\times$3 kernels and a large stride for down-sampling. Then the current scale features are concatenated with all up-sampled and down-sampled features from other scales in the channel dimension, suggesting that each concatenated features can represent the global contextual and local detail information of the lesion. The concatenated features can be directly used for lesion segmentation or RECIST diameter prediction with a convolutional layer of 1 or 4 3$\times$3 kernels. But before that, to further improve the feature representations, we build a scale-aware attention module (SA) based on the channel attention mechanism of DANet \cite{fu2019dual}, which selectively emphasizes interdependent channel features by integrating associated features among all feature channels. The SA's structure is shown in the red box of Fig. \ref{fig:framework}. Different lesions have different scales, but SA is able to adaptively select suitable scale or channel features for them for better accuracy in our tasks. To get a full-size prediction, a deconvolutional layer with 32 4$\times$4 kernels and a stride of 2 is attached to the last SA. Also, 6 and 3 side outputs are added in the decoder to introduce the deep mask supervision (the pink solid arrows) and deep diameter supervision (the pink dotted arrows), respectively. The deep diameter supervision is only used for high-resolution side outputs, because a high-quality RECIST diameter prediction requires large spatial and detailed information.

\textbf{Model Optimization:}
Following \cite{tang2018semi,tang2020one}, we also convert the RECIST diameter prediction problem into a key point regression problem. It means that the model will predict four key point heatmaps to locate the four endpoints of RECIST diameters. For both tasks, a mean squared error loss ($l_{mse}$) is used to compute the errors between predictions and supervisions. As a pixel-wise loss, it will be affected by imbalanced foreground and background pixels. Unfortunately, lesion and non-lesion regions are highly imbalanced at stage 1 of our framework. To deal with this problem, an additional IOU loss \cite{rahman2016optimizing} ($l_{iou}$) is introduced for the lesion segmentation task, which handles the global structures of lesions instead of every single pixel. As described above, 11 side outputs with deep mask supervision and 3 side outputs with deep diameter supervision are used in PDNet. Therefore, the loss is $l_{seg}=\sum_{i=1}^{11}{[l^i_{mse}+l^i_{iou}]}$ for lesion segmentation and $l_{dp}=\sum_{i=1}^{3}{l^i_{mse}}$ for RECIST diameter prediction. The final loss is $l=\lambda l_{seg}+(1-\lambda) l_{dp}$, where $\lambda$ is set to 0.01 to balance the magnitude of the two losses. Two PDNet models used in two stages are trained separately. 

For the segmentation task, we do not have manual lesion masks in DeepLesion. Therefore, we first construct an ellipse from a RECIST annotation following \cite{tang2019uldor}. Then the morphological snake (MS) algorithm \cite{marquez2013morphological} is used to refine the ellipse to get a pseudo mask with good quality, serving as the mask supervision. For $l_{dp}$ update, we generate four 2D Gaussian heatmaps with a standard deviation of $\sigma$ from four endpoints of each RECIST annotation, serving as the diameter supervision. We set $\sigma=3$ at stage 1, $\sigma=7$ at stage 2. We also apply an iterative refining strategy. When the training is done, we run the model over all training data to get their lesion segmentation results, and then use the MS algorithm to refine them. With an ellipse and a refined segmentation result, we can update the pseudo mask by setting their intersections as foreground, their differences as uncertain regions that will be ignored for loss computation during training, and the rest as background. The new pseudo masks can be used to retrain the models. The final models are obtained after three training iterations. 
 
\section{Experiments}

\textbf{Datasets and Evaluation Criteria:}
The DeepLesion dataset \cite{yan2018deeplesion} contains $32,735$ CT lesion images with RECIST diameter annotations from $10,594$ studies of $4,459$ patients. Various lesions throughout the whole body are included, such as lung nodules, bone lesions, liver tumors, enlarged lymph nodes, and so on. Following \cite{cai2018accurate,tang2020one}, $1000$ lesion images from 500 patients with manual segmentations serve as a test set. The rest patient data are used for training.
An external test set with 1350 lesions from 900 patients is built for external validation by collecting lung, liver, pancreas, kidney tumors, and lymph nodes from multiple public datasets, including Decathlon-Lung \cite{simpson2019large} (50), LIDC \cite{armato2011lung} (200), Decathlon-HepaticVessel \cite{simpson2019large} (200), Decathlon-Pancreas \cite{simpson2019large} (200), KiTS \cite{heller2019kits19} (150), and NIH-Lymph Node \cite{roth2014new} (100), specifically. Each lesion has a 3D mask. To make it suitable for evaluation, we select an axial slice for each lesion where the lesion has the largest spatial extent based on its 3D mask. The long and short diameters calculated from the 2D lesion mask of the selected slice are treated as the ground truths of the RECIST diameters.
We utilize the same criteria as in \cite{tang2020one} to compute the quantitative results. The pixel-wise precision, recall, and dice coefficient (Dice) are used for lesion segmentation. The mean and standard deviation of differences between the diameter lengths (mm) of the predictions and manual annotations are used for RECIST diameter prediction.

\textbf{Implementation Details:}
PDNet is implemented in PyTorch. The image encoder is initialized with the ImageNet \cite{DengDSLL009} pre-trained weights. At both stages, we train PDNet using Adam optimizer \cite{kingma2014adam} with an initial learning rate of 0.001 for 120 epochs and decay it by 0.1 after 60 and 90 epochs. During training, all CT images are resized to 512$\times$512 first. Then the input images are generated by randomly rotating by $\theta \in [-10^{\circ}, 10^{\circ}]$ and cropping a square sub-image whose size is $s \in [480, 512]$ at stage 1 and 1.5 to 3.5 times as large as the lesion's long side with random offsets at stage 2. 
They are resized to 512$\times$512 and 256$\times$256 for both stages, respectively.
For testing, input images are generated by resizing to 512$\times$512 at stage 1 or cropping a square sub-image whose size is 2.5 times the long side of lesion segmentation result produced by the first PDNet model at stage 2.
To mimic the clicking behavior of a radiologist, a point is randomly selected from a region obtained by eroding the ellipse to half of its size. Besides requiring a click, it can also cooperate with lesion detection and tracking techniques \cite{tang2019uldor,yan2019mulan,yan2020learning,cai2020deep} to perform fully automatic lesion segmentation and RECIST diameter prediction.
\begin{figure*}[t!]
\centering
\begin{minipage}[b]{\linewidth}
  \begin{minipage}[b]{1.0\linewidth}
		\centering
		\includegraphics[width=0.99\linewidth]{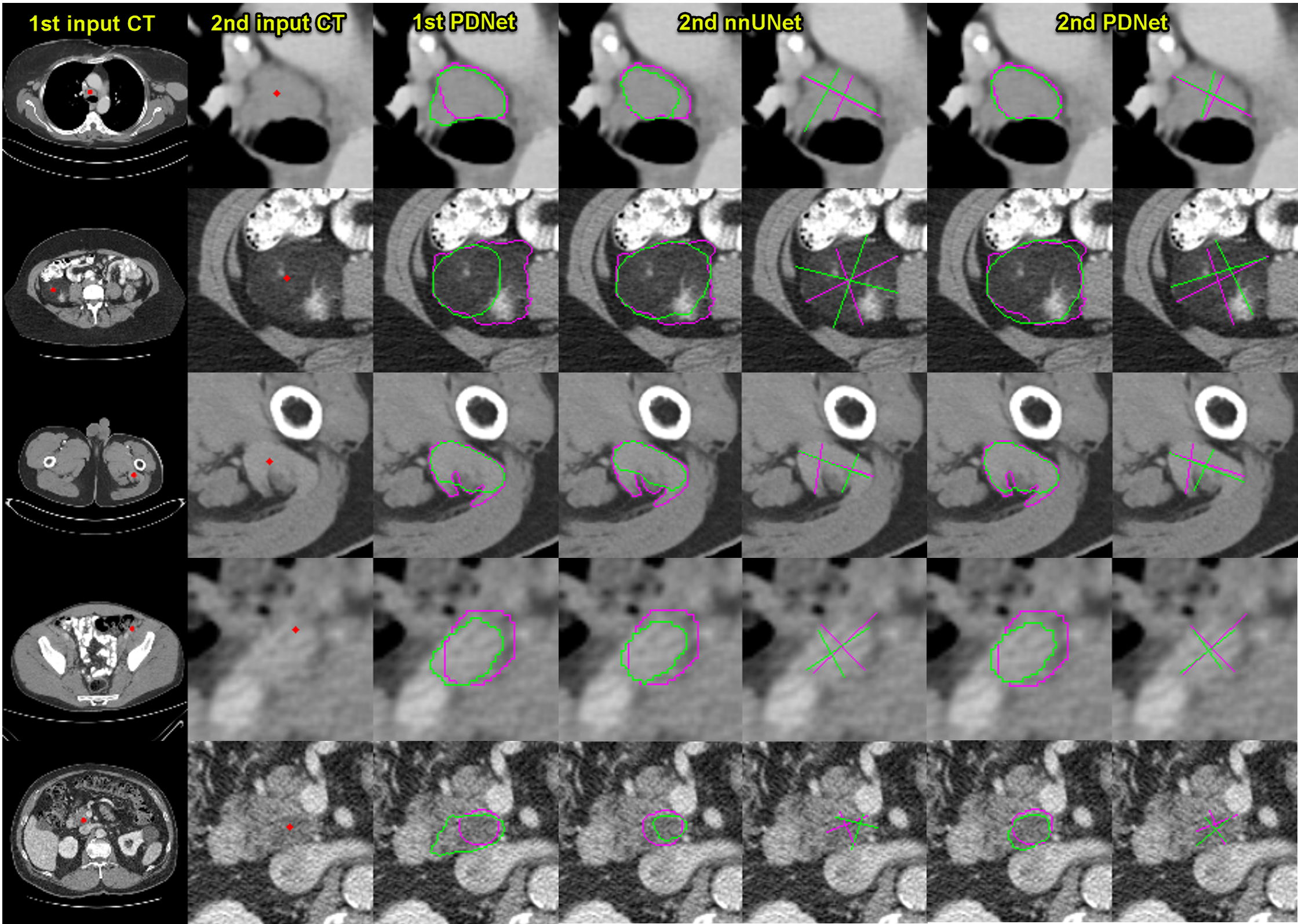} \\
	\end{minipage}
    \begin{minipage}[b]{0.15\linewidth}
		\centering
		\centerline{(a)}\medskip
	\end{minipage}
	\begin{minipage}[b]{0.13\linewidth}
		\centering
		\centerline{(b)}\medskip
	\end{minipage}
	\begin{minipage}[b]{0.13\linewidth}
		\centering
		\centerline{(c)}\medskip
	\end{minipage}
	\begin{minipage}[b]{0.27\linewidth}
		\centering
		\centerline{(d)}\medskip
	\end{minipage}
	\begin{minipage}[b]{0.28\linewidth}
		\centering
		\centerline{(e)}\medskip
	\end{minipage}
\end{minipage}
  \caption{Visual examples of results on the DeepLesion test set (the first three rows) and the external test set (the last two rows), where the pink and green curves/crosses are the manual annotations and automatic results. Given a CT image (a) and a click guidance (red spot), the $1^{st}$ PDNet produces an initial lesion segmentation result (c) at stage 1, based on which a LOI (b) is extracted and taken as input of stage 2. The final results of lesion segmentation (left) and RECIST diameter prediction (right) are obtained by the $2^{nd}$ nnUNet (d) and $2^{nd}$ PDNet (e). Best viewed in color.}
  \label{fig:result}
\end{figure*}

\textbf{Experimental Results:}
As a powerful segmentation framework, nnUNet \cite{isensee2018nnu} built based on UNets \cite{ronneberger2015u} has been successfully used in many medical image segmentation tasks, thus it can serve as a strong baseline in this task. We train an nnUNet model for each stage by taking as input the 3-channel image and using the same setting as PDNet. 
At stage 1, nnUNet produced poor segmentation performance, \eg the DICE score is about 0.857 on the DeepLesion test set, suggesting that the LOI extracted from it is not good enough to serve as the input of stage 2. Therefore, the $2^{nd}$ nnUNet takes as input the LOIs extracted by the $1^{st}$ PDNet, achieving a DICE of 0.911.

Fig. \ref{fig:result} shows five visual examples of the results produced by nnUNet and PDNet. We can see that \textbf{1)} the $1^{st}$ PDNet can segment the lesion region (Fig. \ref{fig:result}(c)), even if they are small (the $4^{th}$ row), heterogeneous (the $2^{nd}$ row), or have blurry boundaries (the $5^{th}$ row), irregular shapes (the $3^{rd}$ row), etc. It indicates that the LOIs can be extracted reliably at stage 1 (Fig. \ref{fig:result}(b)). \textbf{2)} The lesion segmentation results can be improved significantly by the $2^{nd}$ PDNet (Fig. \ref{fig:result}(e)), but a part of them become worse when using the $2^{nd}$ nnUNet (\eg the $1^{st}$ and $3^{rd}$ rows in Fig. \ref{fig:result}(d)). \textbf{3)} The RECIST diameters predicted by PDNet are much closer to the references than nnUNet. 
The qualitative results validate that the proposed framework can segment the lesions and predict their RECIST diameters reliably using only a click guidance. It may struggle when lesions have highly blurry boundaries or irregular shapes (rows 3 and 5), in which all methods will fail to segment them well. 

\begin{table}[t!]
	\begin{center}
		\caption{Results of lesion segmentation and RECIST diameter prediction on two test sets. The mean and standard deviation of all metrics are reported.}
		\label{tab:result}
		{
			\scriptsize
			
			\begin{tabular}{|@{}*{1}{m{2.5cm}<{\centering}@{}}|@{}*{1}{m{1.9cm}<{\centering}@{}|@{}}*{1}{m{1.9cm}<{\centering}@{}|@{}}*{1}{m{1.9cm}<{\centering}@{}|@{}}*{1}{m{1.9cm}<{\centering}@{}|@{}}*{1}{m{1.9cm}<{\centering}@{}|@{}}}
				\hline
				 & \multicolumn{3}{c|}{Lesion segmentation} & \multicolumn{2}{c|}{RECIST diameter prediction} \\ \cline{2-6}
                \multirow{-2}{*}{Method} & Precision & Recall & Dice & Long axis & Short axis \\ \hline
				\multicolumn{6}{|c|}{DeepLesion test set} \\ \hline
				Cai \etal \cite{cai2018accurate}  &  0.893$\pm$0.111  &  0.933$\pm$0.095 & 0.906$\pm$0.089  & -  & -  \\ 
				Tang \etal \cite{tang2018semi}  &  - & - & - & 1.893$\pm$2.185 &  1.614$\pm$1.874  \\ 
				Tang \etal \cite{tang2020one}  & 0.883$\pm$0.057  &  \textbf{0.947$\pm$0.074} &  0.912$\pm$0.039 &  1.747$\pm$1.983 & 1.555$\pm$1.808   \\ 
				nnUNet \cite{isensee2018nnu}  & \textbf{0.977$\pm$0.033} & 0.852$\pm$0.086 & 0.907$\pm$0.050 &  2.108$\pm$1.997 & 1.839$\pm$1.733   \\
				PDNet  & 0.961$\pm$0.044  &  0.898$\pm$0.077 &  \textbf{0.924$\pm$0.045} &  \textbf{1.733$\pm$1.470} & \textbf{1.524$\pm$1.374}   \\ \hline
				\multicolumn{6}{|c|}{External test set} \\ \hline
				nnUNet \cite{isensee2018nnu}  & \textbf{0.946$\pm$0.062} & 0.815$\pm$0.099 & 0.870$\pm$0.054 &  2.334$\pm$1.906 & 1.985$\pm$1.644   \\
				PDNet  & 0.927$\pm$0.074  &  \textbf{0.857$\pm$0.093} &  \textbf{0.885$\pm$0.049} &  \textbf{2.174$\pm$1.437} & \textbf{1.829$\pm$1.339}   \\ \hline
			\end{tabular}
		}
	\end{center}
\end{table}

\begin{table}[!t]
    \begin{center}
    {
		\caption{Category-wise results in terms of segmentation Dice and the prediction error of diameter lengths on the external test set.}
		\label{tab:category-result}
		\scriptsize
        \begin{tabu} to 0.99\textwidth {| X[c] | X[c] | X[c] | X[c] | X[c] | X[c] |}
              \hline
            Method & Lung & Liver  &  Pancreas & Kidney  &  Lymph node  \\ \hline
            \multicolumn{6}{|c|}{Lesion segmentation (Dice)} \\ \hline
        	nnUNet \cite{isensee2018nnu} & 0.853$\pm$0.054  &  0.876$\pm$0.057  &  0.877$\pm$0.055 & 0.890$\pm$0.057  &  0.865$\pm$0.050  \\ 
        	PDNet & 0.876$\pm$0.046 &  0.893$\pm$0.051  &  0.886$\pm$0.050 & 0.911$\pm$0.050  &  0.876$\pm$0.045  \\
             \hline
            \multicolumn{6}{|c|}{RECIST diameter prediction (long axis)} \\ \hline
        	nnUNet \cite{isensee2018nnu} & 2.396$\pm$2.004 &  2.862$\pm$2.090   &  2.655$\pm$2.048 & 2.493$\pm$1.963  &  1.958$\pm$1.639  \\ 
        	PDNet & 2.435$\pm$1.461 &  2.378$\pm$1.463  &  2.220$\pm$1.536 & 2.603$\pm$1.533  &  1.897$\pm$1.293  \\
             \hline
            \multicolumn{6}{|c|}{RECIST diameter prediction (short axis)} \\ \hline
        	nnUNet \cite{isensee2018nnu} & 2.223$\pm$1.404  &  2.383$\pm$1.808  &  2.242$\pm$1.637 & 2.342$\pm$1.854  &  1.712$\pm$1.440   \\ 
        	PDNet & 2.243$\pm$1.333  &  2.168$\pm$1.405   &  1.977$\pm$1.359 & 2.362$\pm$1.488   &  1.486$\pm$1.174   \\
             \hline
             
        \end{tabu}
    }
    \end{center}
\end{table}%

Table \ref{tab:result} lists the quantitative results of different methods on two test sets. It can be seen that \textbf{1)} compared to the best previous work \cite{tang2020one}, PDNet boosts the Dice score by a large margin of 1.2\% (from 0.912 to 0.924), and also gets smaller diameter errors on the DeepLesion test set. It means that PDNet can simultaneously segment the lesions accurately and produce reliable RECIST diameters close to the radiologists' manual annotations. \textbf{2)} Compared to the strong baseline nnUNet, PDNet gets much better results on both test sets. This is because PDNet is able to extract more comprehensive multi-scale features to better represent the appearances of different kinds of lesions. \textbf{3)} Compared to the DeepLesion test set, the performance drops for both nnUNet and PDNet on the external test set, \eg the Dice score of PDNet decreases from 0.924 to 0.885. In the external test set, some lesion masks are not well annotated, thus the generated ground-truth RECIST diameters will also be affected.
The $4^{th}$ row of Fig. \ref{fig:result} shows an unsatisfactory annotation, where the manual annotation is larger than the lesion's actual size. Meanwhile, the segmentation results produced by PDNet are better aligned to the lesion boundaries visually.

Table \ref{tab:category-result} lists the category-wise results on the external test set. PDNet achieves better performance in terms of all metrics and categories except the RECIST diameter prediction on lung and kidney tumors. After investigation, a potential reason is that a part of lung and kidney tumors have highly irregular shapes, whose diameters generated from manual masks are very likely to be larger
, and nnUNet tends to predict larger diameters than PDNet in these cases. These results evidently demonstrate the effectiveness and robustness of our method.

\textbf{Ablation Studies:}
To investigate the contributions of PDNet's components, \ie prior encoder (PE), top-down connection (T2D), bottom-up connection (B2U), and scale-aware attention module (SA), we configure different models by sequentially adding them into the base model that includes the image encoder with input of the 3-channel image and a UNet-style decoder. Table \ref{tab:ablation} outlines their quantitative comparisons. As can be seen, \textbf{1)} each added component improves the performance at both stages, demonstrating that the proposed strategies contribute to learning more comprehensive features for our tasks. \textbf{2)} The largest improvement gain is brought by introducing PE, especially for stage 1, demonstrating that PE can effectively explore the click prior information to learn lesion-specific attention matrices which heavily enhances the extracted multi-scale features for performance improvement.

\begin{table}[!t]
    \begin{center}
    {
		\caption{Results of different settings of our method in terms of Dice and the prediction error of diameter lengths on the DeepLesion test set.}
		\label{tab:ablation}
		\scriptsize
        \begin{tabu} to 0.99\textwidth {| X[0.3c] | X[0.3c] | X[0.3c] | X[0.3c] | X[0.3c] | X[c] | X[c] | X[c] | X[c] |}
              \hline
              \multicolumn{5}{|c|}{Settings} & \multicolumn{1}{c|}{Stage 1} & \multicolumn{3}{c|}{Stage 2} \\ \cline{1-9}
                \multirow{6}{*}{\rotatebox[origin=c]{270}{Base model}} & PE & T2D & B2U & SA & Dice & Dice & Long axis & Short axis\\ \cline{2-9}
              &  &  &  &  & 0.871$\pm$0.123  &  0.909$\pm$0.068 & 1.961$\pm$2.278  &  1.704$\pm$1.948  \\ 
        	  & \checkmark &  &  &  &  0.890$\pm$0.089  &  0.915$\pm$0.055 & 1.861$\pm$1.934  &  1.617$\pm$1.684  \\ 
        	  & \checkmark & \checkmark &  &  &  0.900$\pm$0.067  &  0.919$\pm$0.054 & 1.809$\pm$1.731  &  1.577$\pm$1.508  \\ 
        	  & \checkmark & \checkmark & \checkmark &  &  0.905$\pm$0.070  &  0.921$\pm$0.050 & 1.758$\pm$1.696  &  1.544$\pm$1.470  \\ 
        	  & \checkmark & \checkmark & \checkmark & \checkmark &  0.911$\pm$0.060  &  0.924$\pm$0.045 & 1.733$\pm$1.470 & 1.524$\pm$1.374  \\
              \hline
        \end{tabu}
    }
    \end{center}
\end{table}%

\section{Conclusions}
This paper proposes a novel deep neural network architecture, prior-guided dual-path network (PDNet), for accurate lesion segmentation and RECIST diameter prediction. It works in a two-stage manner. Providing very simple human guide information, an LOI can be extracted precisely by segmentation at stage 1 and its segmentation and RECIST diameters can be predicted accurately at stage 2. As such, it offers a useful tool for radiologists to get reliable lesion size measurements (segmentation and RECIST diameters) with greatly reduced time and labor. It can potentially provide high positive clinical values.

\bibliographystyle{ieeetr}
\bibliography{egbib}
\end{document}